\documentclass[journal]{IEEEtran}
\IEEEoverridecommandlockouts
\usepackage{booktabs}
\usepackage{siunitx}
\usepackage{cite}
\usepackage{amsmath,amssymb,amsfonts}
\usepackage{graphics, framed}
\usepackage{graphicx, enumerate}
\usepackage{epsfig}
\usepackage{epstopdf}
\usepackage{url}
\usepackage{multimedia}
\usepackage{paralist}
\usepackage[printonlyused]{acronym}
\usepackage{units}
\usepackage{algorithmic}
\usepackage{textcomp}
\usepackage{float}
\usepackage{tabularx}
\usepackage{balance}
\usepackage{subfigure}
\usepackage{xcolor}
\def\BibTeX{{\rm B\kern-.05em{\sc i\kern-.025em b}\kern-.08em
T\kern-.1667em\lower.7ex\hbox{E}\kern-.125emX}}
\usepackage{mathtools}
\usepackage{cuted} 
\usepackage[export]{adjustbox}

\usepackage{amsmath}
\usepackage{amssymb}
\usepackage{dblfloatfix}
\usepackage{lipsum}
\usepackage{color,soul}
\usepackage{multirow}
\usepackage[overload]{textcase}
\newcommand{\FGR}[1]{Fig.~\ref{#1}}

\newcommand{\SEC}[1]{Section~\ref{#1}}

\newcommand{\minus}{\scalebox{0.75}[1.0]{$-$}}
\acrodef{5G}[5G]{5\textsuperscript{th}-Generation}
\acrodef{BW}[BW]{bandwidth}
\acrodef{BER}[BER]{bit error rate}
\acrodef{BPSK}[BPSK]{binary phase-shift keying}
\acrodef{CW}[CW]{continuous wave}
\acrodef{CSI}[CSI]{channel state information}
\acrodef{D2D}[D2D]{device-to-device}
\acrodef{dB}[dB]{decibel}
\acrodef{dBi}[dBi]{decibel isotropic}
\acrodef{dBm}[dBm]{decibel over a milliwatt}
\acrodef{DSN}[DSN]{deep space network}
\acrodef{DTN}[DTN]{delay tolerant network}
\acrodef{Gbps}[Gbps]{gigabit per second}
\acrodef{GHz}[GHz]{gigahertz}
\acrodef{GAT}[GAT]{graph attention network}
\acrodef{THz}[THz]{Terahertz}
\acrodef{ISL}[ISL]{inter-satellite link}
\acrodef{RIS}[RIS]{reconfigurable intelligent surface}
\acrodef{GM}[GM]{Gamma mixture}
\acrodef{PSK}[PSK]{phase shift keying}
\acrodef{QAM}[QAM]{quadrature amplitude modulation}
\acrodef{AWGN}[AWGN]{additive white Gaussian noise}
\acrodef{SNR}[SNR]{signal-to-noise ratio}
\acrodef{AF}[AF]{amplitude-and-forward}
\acrodef{MIMO}[MIMO]{multiple-input multiple-output}
\acrodef{mMIMO}[mMIMO]{massive-multiple-input multiple-output}
\acrodef{SDN}[SDN]{Software-defined network}
\acrodef{SON}[SON]{self-organizing network}
\acrodef{HetNet}[HetNet]{heterogeneous network}
\acrodef{FSO}[FSO]{free-space optics}
\acrodef{UM-MIMO}[UM-MIMO]{ultra-massive-MIMO}
\acrodef{AP}[AP]{access point}
\acrodef{UE}[UE]{user equipment}
\acrodef{NTN}[NTN]{non-terrestrial network}
\acrodef{UAV}[UAV]{unmanned aerial vehicle}
\acrodef{HAPS}[HAPS]{high-altitude platform station}
\acrodef{LEO}[LEO]{low-Earth orbit}
\acrodef{BAN}[BAN]{body area network}
\acrodef{WLAN}[WLAN]{wireless local area network}
\acrodef{QoS}[QoS]{quality of service}
\acrodef{TCS}[TCS]{thermal control system}
\acrodef{QCL}[QCL]{quantum cascade laser}
\acrodef{CMOS}[CMOS]{complementary metal-oxide semiconductor}
\acrodef{V-HetNet}[V-HetNet]{vertical heterogeneous network}
\acrodef{DL}[DL]{Deep learning}
\acrodef{DRL}[DRL]{deep reinforcement learning}
\acrodef{EIRP}[EIRP]{effective isotropic radiated power}
\acrodef{FDTD}[FDTD]{Finite-difference time-domain}
\acrodef{FEM}[FEM]{finite element method}
\acrodef{MoM}[MoM]{method of moments}
\acrodef{VNA}[VNA]{vector network analyzer}
\acrodef{CS}[CS]{channel sounder}
\acrodef{CIR}[CIR]{channel impulse response}
\acrodef{CTF}[CTF]{channel transfer function}
\acrodef{DPM}[DPM]{Dirichlet process mixture}
\acrodef{TOA}[TOA]{time of arrival}
\acrodef{GNN}[GNN]{graph neural network}
\acrodef{IoT}[IoT]{Internet of Things}
\acrodef{MLE}[MLE]{maximum likelihood estimation}
\acrodef{LOS}[LOS]{line-of-sight}
\acrodef{NLOS}[NLOS]{non-line-of-sight}
\acrodef{SG}[SG]{signal generator}
\acrodef{SEP}[SEP]{Sun-Earth-probe}
\acrodef{FDSOI}[FDSOI]{fully depleted silicon on insulator}
\acrodef{OpEx}[OpEx]{operational expenditures}
\acrodef{TCO}[TCO]{total cost of ownership}
\acrodef{CapEx}[CapEx]{capital expenditures}
\acrodef{MAC}[MAC]{medium access control}
\acrodef{GEO}[GEO]{geostationary orbit}
\acrodef{SWaP}[SWaP]{size, weight, and power}
\acrodef{NOMA}[NOMA]{Non-orthogonal multiple access}
\acrodef{GAL}[GAL]{graph attention layer}
\acrodef{RMSE}[RMSE]{root mean square error}
\acrodef{MAE}[MAE]{Mean absolute error}

\ifCLASSINFOpdf
\else
\fi

\begin{document}
\title{Graph Attention Network-Based Single-Pixel Compressive Direction of Arrival Estimation}

\author{K{\"{u}}r{\c{s}}at~Tekb{\i}y{\i}k,~\IEEEmembership{Graduate Student Member,~IEEE,} Okan~Yurduseven,~\IEEEmembership{Senior~Member,~IEEE}, G{\"{u}}ne{\c{s}}~Karabulut~Kurt,~\IEEEmembership{Senior~Member,~IEEE}

\thanks{Manuscript received xx; revised xx; accepted xx. The work of O. Yurduseven was supported by a research grant from
the Leverhulme Trust under the Research Leadership Award RL-2019-019.} 
\thanks{K. Tekb{\i}y{\i}k is with the Department of Electronics and Communications Engineering, {\.{I}}stanbul Technical University, {\.{I}}stanbul, Turkey (e-mail: tekbiyik@itu.edu.tr).}
\thanks{O. Yurduseven is with the Centre for Wireless Innovation, Queen's University, Belfast, Belfast BT3 9DT, U.K. (e-mail: okan.yurduseven@qub.ac.uk).}
\thanks{G. Karabulut Kurt is with the Department of Electrical Engineering, Polytechnique Montr\'eal, Montr\'eal, Canada (e-mail: gunes.kurt@polymtl.ca).}

}

\IEEEoverridecommandlockouts 

\maketitle

\begin{abstract}
In this paper, we present a single-pixel compressive direction of arrival (DoA) estimation technique leveraging a graph attention network (GAT)-based deep-learning framework. The physical layer compression is achieved using a coded-aperture technique, probing the spectrum of far-field sources that are incident on the aperture using a set of spatio-temporally incoherent modes. This information is then encoded and compressed into the channel of the coded-aperture. \textcolor{black}{The coded-aperture is based on a metasurface antenna design and it works as a receiver, exhibiting a single-channel and replacing the conventional multi-channel raster scan-based solutions for DoA estimation.} The GAT network enables the compressive DoA estimation framework to learn the DoA information directly from the measurements acquired using the coded-aperture. This step eliminates the need for an additional reconstruction step and significantly simplifies the processing layer to achieve DoA estimation. We show that the presented GAT integrated single-pixel radar framework can retrieve high fidelity DoA information even under relatively low signal-to-noise ratio (SNR) levels. 
\end{abstract}

\IEEEpeerreviewmaketitle
\acresetall

\section{Introduction}\label{sec:intro}
Direction of arrival (DoA) estimation has been the subject of much recent research, particularly in the context of channel characterization for wireless communications~\cite{guan2019channel,ma2020impact}. Conventional receiver architectures \textcolor{black}{for} DoA estimation leverage antenna array-based solutions with a dedicated signal processing layer, such as MUSIC~\cite{chen2018two}, ESPRIT~\cite{wang2016non}, and SAGE~\cite{wu2021measurement}. \textcolor{black}{Antenna arrays typically have elements separated by $\lambda/2$ where $\lambda$ is free-space wavelength.} As a result, array-based topologies require that the received signal is collected through each antenna element within the array. Such an approach can significantly increase the hardware complexity due to an excessive number of data acquisition channels, particularly when the electrical size of the aperture is large and the operating frequency is increased. 
Recently, the concept of compressive sensing has received significant \textcolor{black}{attraction} as an enabling technology \textcolor{black}{for DoA estimation \cite{10.1145/3412060.3418432,domae2021machine}. In \cite{10.1145/3412060.3418432,domae2021machine}, the compressive DoA estimation study was facilitated using a phased array aperture. The phased array aperture requires that each array element has a dedicated phase shifting circuit. As an alternative approach, single-pixel, wave-chaotic metasurface antennas have recently been shown to offer significant potential for compressive sensing, particularly in millimeter-wave (mmW) imaging~\cite{imani2020review}. A significant advantage of the single-pixel coded-aperture concept is that these apertures do not require a dedicated feeding network to control the excitation phase of each array antenna element}. \textcolor{black}{These metasurface apertures are called wave-chaotic since they can radiate spatially-varying, quasi-random radiation patterns over time. These radiation patterns are called spatio-temporally incoherent because of the low level of correlation that exists between the radiation patterns synthesized by the metasurface aperture.} The underlying principle behind this concept is that, the scene information can be encoded onto spatio-temporally incoherent wave-chaotic modes radiated by single-pixel compressive coded-aperture antennas. \textcolor{black}{It has been proven that such an approach can substantially simplify the physical layer architecture \cite{imani2020review}.} 

Leveraging the fundamentals of the compressive sensing concept, the authors previously developed a compressive DoA estimation technique using a single-pixel coded-aperture~\cite{yurduseven2019frequency} and a \textcolor{black}{sparse array-based} receiver topology~\cite{karabulut2004angle}. The single-pixel technique can successfully retrieve the DoA information of sources arbitrarily defined in the far-field of the aperture using only a single channel to acquire the data. Despite the success of these proof-of-concept studies and encouraging results obtained in these works, compressive DoA estimation concept currently suffers from two limitations: First, a priori knowledge of the fields radiated by the compressive antenna must be known. This suggests that a characterization step is needed to obtain the transfer function of the antenna. Second, the compressed signal received at the antenna channel needs to be interacted with the transfer function of the antenna to recover the DoA pattern. This suggests that a reconstruction step is needed. Both these steps are computationally expensive and pose a significant challenge for practical applications. The main motivation of this work is to develop a single-pixel DoA estimation technique leveraging a deep-learning layer to directly estimate the DoA information from the compressed channel measurements, eliminating the need for the reconstruction step altogether. To this end, we develop a Graph Attention Network (GAT) approach embedded within a coded-aperture-based, single-pixel radar framework for DoA estimation.
The contributions of the presented study are summarized as follows:
\begin{enumerate}[{C}1]
\item {We propose a compressive, single-pixel framework that can significantly simplify the receiver physical layer architecture for DoA estimation.} 
\item {We present the first deep learning-based approach as applied to \textcolor{black}{coded-aperture-based} compressed channel measurements for DoA estimation. The proposed method is applied directly to the raw data measured at the compressed channel of the single-pixel receiver.} 
\item {We analyze the performance of the GAT integrated single-pixel compressive DoA estimation technique under various signal-to-noise ratio (SNR) level conditions to present the potential of this technique, \textcolor{black}{as a proof-of-concept}, for channel characterization.} 
\end{enumerate}

The outline of this paper is as follows: In Section \ref{sec:preliminaries}, we explain GAT and the concept of compressive sensing as applied to DoA estimation. In Section \ref{sec:method}, we present the deep learning-based single-pixel DoA estimation scheme integrated with GATs as an enabling learning approach. Finally, in Section \ref{sec:conclusion}, we provide the concluding remarks. 

\section{Preliminaries}\label{sec:preliminaries}

\subsection{Graph Attention Networks}\label{sec:gat}

\textcolor{black}{Conventional deep learning methods such as CNN can provide accurate results on challenging tasks.} However, they suffer from data without grid-like structure~\textcolor{black}{\cite{velickovic_graph_2018}}. As a state-of-the-art method, \acp{GNN} have been recently proposed to deal with data without grid-like structure~\textcolor{black}{\cite{brody2021attentive}}. By using graph structure and node features, \acp{GNN} can learn the representations of nodes and the graph. Furthermore, the composition of \ac{GNN} with attention mechanism creates a more robust method, namely \acp{GAT}, with inductive learning capability~\textcolor{black}{\cite{velickovic_graph_2018}}. Therefore, it is expected that \acp{GAT} can be utilized for DoA problems due to the random position of the source and random propagation environment. 

The fundamental part of a \ac{GAT} is \ac{GAL} with $K$ input nodes. The input nodes are defined as follows:
\textcolor{black}{\begin{align}
    \mathbf{\gamma}=\left\{\vec{\gamma}_{1}, \vec{\gamma}_{2}, \ldots, \vec{\gamma}_{K}\right\}, \, \vec{\gamma}_{i} \in \mathbb{R}^{F},
\end{align}}
where $F$ denotes the number of features in each node. At the output of \ac{GAL}, the cardinality of the output set might differ from the input set. Therefore, let us denote the number of features of nodes as $F^{\prime}$ without loss of generality. In each \ac{GAL}, a linear operation that is defined by a weight matrix, \textcolor{black}{$\mathbf{W} \in \mathbb{R}^{F^{\prime}\times F}$}, transforms the input attributes of nodes to higher-level attributes. Following the linear transformation, self-attention on nodes is performed by the self-attention mechanism, $a: \mathbb{R}^{F^{\prime}} \times \mathbb{R}^{F^{\prime}} \rightarrow \mathbb{R}$. The attention mechanism computes attention coefficients \textcolor{black}{according to}:
\begin{align}
c_{i j}=a\left(\mathbf{W} \vec{\gamma}_{i}, \mathbf{W} \vec{\gamma}_{j}\right),
\end{align}
where the neighborhood between the $i$-th and $j$-th nodes is shown by $c_{i j}$. In other words, the attention coefficient shows how much of an effect the attributes of the $j$-th node have on the $i$-th node. By using the softmax function, we normalize the coefficients as follows:
\begin{align}
\alpha_{i j}=\operatorname{softmax}_{j}\left(c_{i j}\right)=\frac{\exp \left(c_{i j}\right)}{\sum_{k \in \mathcal{N}_{i}} \exp \left(c_{i k}\right)},
\end{align}
where $\mathcal{N}_{i}$ denoted neighborhood of $i$-th node, which is defined by a binary adjacency matrix $\mathbf{A} \in\{0,1\}^{K \times K}$. Then, the convolution operation is performed over \textcolor{black}{the} graph\textcolor{black}{, and modeled as:}
\begin{align}
\mathbf{Z}=\alpha \mathbf{X} \mathbf{W}+\mathbf{b},
\end{align}
where $\mathbf{X} \in \mathbb{R}^{K \times F^{\prime}}$ and $\mathbf{b}$ are the node attributes matrix and the trainable bias vector, respectively. Finally, the graph is reduced to a single vector by global pooling layer~\cite{lee_self-attention_2019}. It enables generalization of the network and decreases the number of representations. Thus, global pooling layer can avoid the network from overfitting. For more information on \ac{GAT} and its previous application on channel estimation, we refer the readers to~\cite{tekbiyik2021channel}.

\subsection{Compressive Sensing and DoA Estimation}\label{sec:compressing_sensing}

Compressive sensing is a computational technique that can be used to solve a sparse problem from a reduced number of measurements~\cite{donoho2006compressed}. In \FGR{fig:System_Depiction}, we demonstrate the compressive aperture concept facilitated for the DoA estimation problem. \textcolor{black}{In the presented concept, the coded-aperture works as a receiver whereas the far-field source works as a transmitter. It should be noted that, if implemented in a wireless communications scenario, such as channel characterization, the proposed technique can be considered suitable for implementation at the network side to identify the DoA information of far-field sources that are incident on the coded-aperture.}
\begin{figure}[!t]
    \centering
    \includegraphics[width=\linewidth]{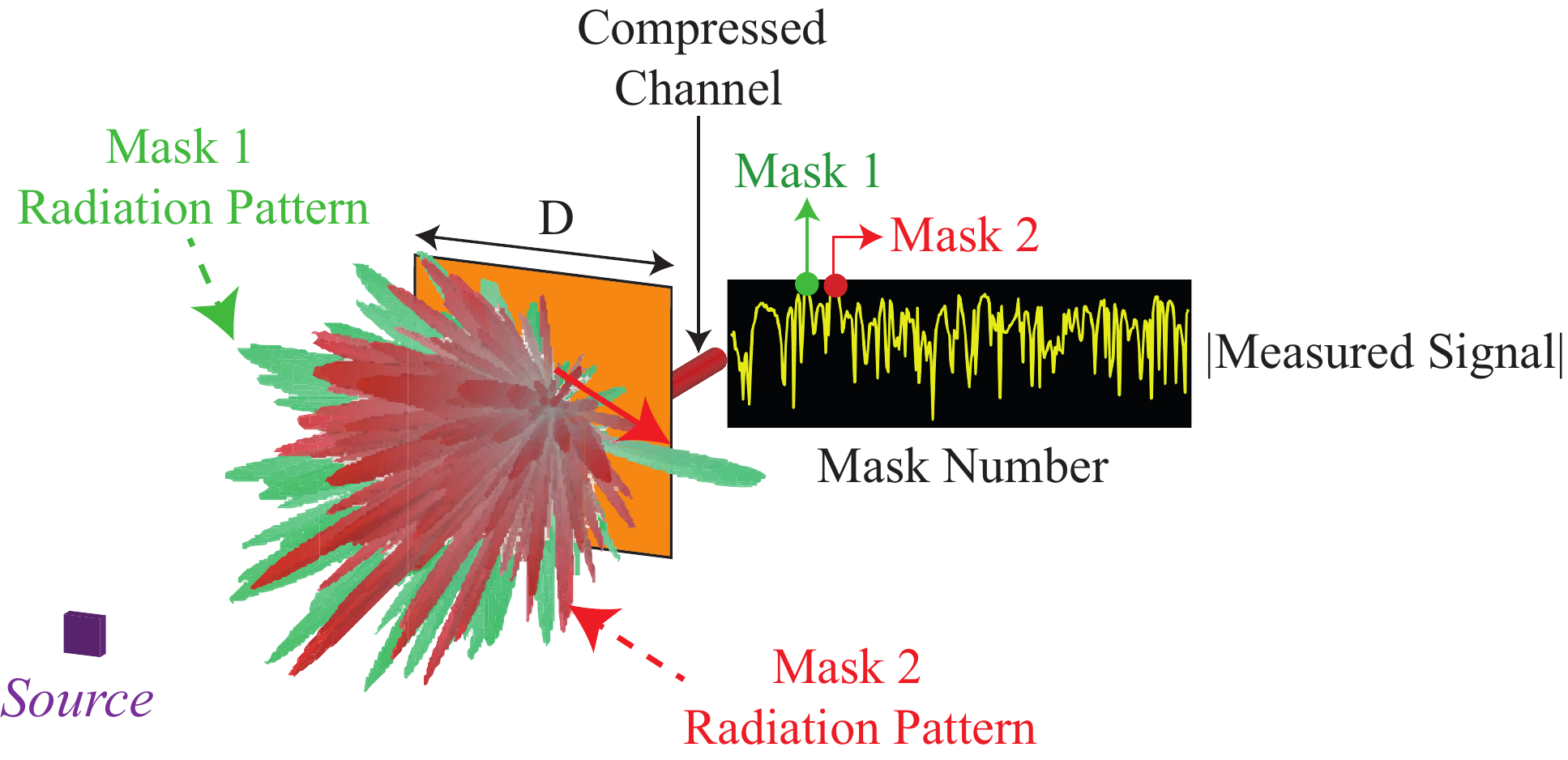}
    \caption{\textcolor{black}{Depiction of the coded-aperture-based compressive DoA estimation technique. Whereas the measured signal is shown for all 2000 masks, the values for mask 1 and mask 2 are highlighted for this depiction. $D=25$ cm.}}
    \label{fig:System_Depiction}
\end{figure}

\begin{figure*}[!t]
    \centering
    \includegraphics[width=\linewidth]{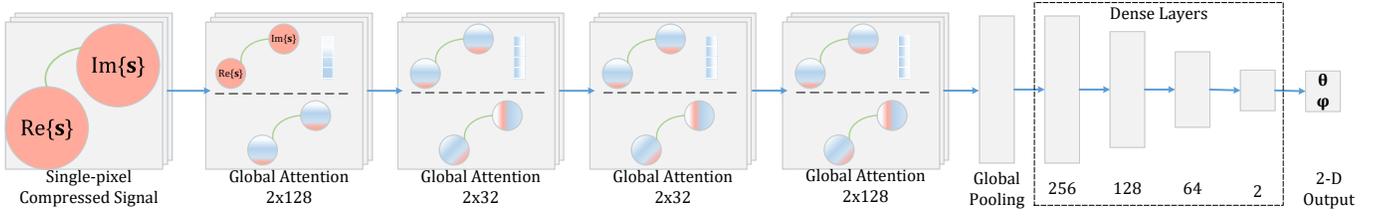}
    \caption{\textcolor{black}{The proposed \ac{GAT} consists of four \acp{GAL} and four dense layers respectively for creating feature maps over graph-structured data and regression analysis between features and 2-D angle values.}}
    \label{fig:gat_scheme}
\end{figure*}

The size of the synthesized aperture is $D=25$ cm and the operating frequency is chosen to be 28 GHz. The compressive aperture depicted in \FGR{fig:System_Depiction} is a coded-aperture, suggesting that it radiates spatio-temporally incoherent radiation patterns by means of reconfiguring the aperture~\cite{imani2020review}. In this concept, each aperture reconfiguration state can be considered a \textit{mask} with each mask radiating a different radiation pattern. \textcolor{black}{A careful investigation of the wave-chaotic radiation patterns depicted in \FGR{fig:System_Depiction} reveals that the radiation pattern for each mask can be considered \textcolor{black}{as} a collection of sidelobes, rather than a well-defined main-lobe with a fixed 3dB beamwidth.} 

For the compressive DoA system depicted in \FGR{fig:System_Depiction}, we use a total number of 2000 masks. \textcolor{black}{It should be noted that the dynamic modulation of the coded-aperture can be realized using transistors \cite{imani2020review,smith2017analysis}. These elements exhibit an extremely fast switching response, typically on the order of a few nanoseconds. As a result, sweeping through 2000 masks can be achieved over a microsecond time-frame. This is smaller than the typical coherence time for 5G channels. The selection of the masks is done on a random basis.} 

As shown in \FGR{fig:System_Depiction}, the aperture has a single channel for data acquisition.
A significant advantage of the physical layer compression depicted in~\FGR{fig:System_Depiction} can be appreciated when considering the same size aperture synthesized using an array topology. At the conventional $\lambda/2$ limit, 0.54 cm, the aperture in \FGR{fig:System_Depiction} would require 2209 elements within the array.
Considering the compressive DoA estimation scenario presented in \FGR{fig:System_Depiction}, the measured signal at the compressed channel can be represented as follows:
\begin{equation}\textcolor{black}{
     \mathbf{s}_{M\times1}=\mathbf{E}_{M\times N}\mathbf{P}_{N\times1}+\mathbf{n}_{M\times1}}.
    \label{Equation_6}
\end{equation}

Eq. (\ref{Equation_6}) is known as the forward-model. In Eq. (\ref{Equation_6}), \textbf{s} represents the acquired data at the compressed channel, \textbf{E} is the radiated field (or the transfer function) of the coded-aperture, \textbf{P} is the projection of the far-field source (or sources) on the aperture of the antenna and \textbf{n} is the noise, which controls the SNR level in the acquired data. It should be mentioned that the bold font here denotes the vector-matrix notation. In Eq. (\ref{Equation_6}), $M$ and $N$ denote the number of masks and \textcolor{black}{number of pixels in the projection of the far-field source on the antenna aperture}, respectively. \textcolor{black}{In this context, a further insight into the \textbf{E} and \textbf{P} is needed to appreciate the physical layer compression concept. The transfer function of the antenna across a plane in front of its aperture can be calculated as \cite{lipworth2015comprehensive}:} 
\begin{align}\textcolor{black}{
\mathbf{E(\mathbf{r})}=i\frac{\omega \mu_{0}}{4\pi}{\sum_{n}(\mathbf{m}_{n}\times\mathbf{r})}\Bigg(\frac{ik}{R_{n}}-\frac{1}{R_{n}^2}\Bigg)e^{jkR_{n}}},
\label{Equation_E}
\end{align}
\textcolor{black}{where $\bf{r}$ denotes the coordinates across the plane defined in front of the coded-aperture, $R_{n}=|\bf{r}-\bf{\rho}_{n}|$, and $\bf{\rho}_{n}$ and $\bf{m}_{n}$ denote the coordinates and the magnetic dipole of the $n^{th}$ metamaterial element in the metasurface layer. Similarly, the projection of the far-field source across the same plane in front of the coded-aperture is calculated as follows \cite{yurduseven2019frequency}:} 
\begin{align}\textcolor{black}{
\mathbf{P(\mathbf{r})}=e^{-jk_{0}(y\sin{\theta}\cos{\varphi})(z\sin{\theta}\sin{\varphi})}},
\label{Equation_P}
\end{align}

\begin{figure}[!t]
    \centering
    \includegraphics[width=8cm]{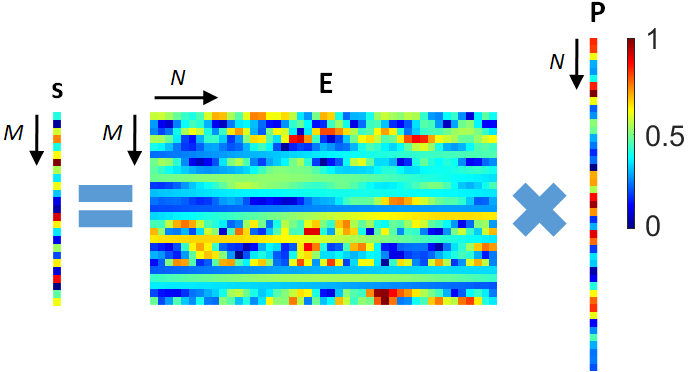}
    \caption{\textcolor{black}{Visualization of the compressive DoA estimation framework: \textbf{P} is sampled by \textbf{E} generating \textbf{s}. For this depiction, $M=25$ and $N=50$, and normalized linear amplitude is shown.}}
    \label{fig:s_E_P}
\end{figure}

\textcolor{black}{In Eq. (\ref{Equation_P}), $\theta$ and $\varphi$ are the incident angles of the far-field source while $k_0$ is the wavenumber. In order to visualize the relationship between \textbf{s}, \textbf{E} and \textbf{P}, in Fig. \ref{fig:s_E_P}, we provide a depiction of the mathematical model for the compressive DoA estimation concept. As depicted in Fig. \ref{fig:s_E_P}, the mapping of $\bf{P}$ onto $\bf{s}$ is achieved through the transfer function of the coded-aperture, $\bf{E}$.}

From Eq. (\ref{Equation_6}), by performing a phase compensation on the transfer function and applying the compensated transfer function to the compressed channel data, an estimate of the field projection pattern, $\textbf{P}_{est}$, can be retrieved. To achieve this, several computational techniques, such as the least-squares method minimizing the objective function $||\mathbf{E}^\dagger\mathbf{s}-\mathbf{P}||_2^2$, where $.^\dagger$ denotes the phase conjugation operator and $||.||_2$ denotes the Euclidean operator, can be used. 

Analyzing Eq. (\ref{Equation_6}), it can be seen that there are two \textcolor{black}{fundamental} challenges with the compressive DoA estimation technique. First, it is evident that the signal captured at the coded-aperture antenna channel is correlated to the far-field source through the transfer function of the antenna. As a result, achieving the DoA estimation using the compressive coded-aperture technique requires a \textcolor{black}{priori} knowledge of the antenna transfer function. This step makes it necessary to measure the radiated fields from the coded-aperture as part of a calibration step, requiring a hardware-intense, highly complex characterization process~\cite{sleasman2017near}. Moreover, in characterizing the transfer function of the antenna, an excellent phase accuracy is needed. \textcolor{black}{This is because} any errors introduced to the transfer function of the antenna during the characterization process can significantly distort the forward-model in Eq. (\ref{Equation_6})~\cite{yurduseven2018relaxation}. Second, the DoA estimation requires that Eq. (\ref{Equation_6}) \textcolor{black}{is} solved to recover $\textbf{P}_{est}$. This reconstruction step can take a considerable amount of time, which is of particular concern for applications where the DoA estimation needs to be done in real-time. Alternatively, retrieving the DoA information directly from the compressed channel measurements can eliminate the \textcolor{black}{need to know the antenna transfer function} and the necessity for the reconstruction step. 

\section{Single-Pixel 2-D DoA with GATs}\label{sec:method}

To address the challenges of the compressive DoA estimation concept, it is desirable that an alternative technique is developed that can work with the channel data without the need to process the forward-model of Eq. (\ref{Equation_6}). Such a technique can eliminate the need to consider the DoA estimation problem as a reconstruction problem. Therefore, the ultimate goal of this work is to develop a \textcolor{black}{GAT-based} deep learning solution that can directly learn from the compressed channel data.

GAT has been recently proposed for learning over graph-structural data by exploiting the self-attention mechanism. In this study, GATs are utilized to deduce the relation between compressed channel data and DoA of the received signal. 

First, we converted the received compressive signal into graph structural data given with $\mathbf{X}$ by mapping the imaginary and real parts of the signal into the attributes of graph nodes as follows:
\textcolor{black}{\begin{align}
\mathbf{X} =[\operatorname{Re}\{\mathbf{s}^\top\} ; \operatorname{Im}\{\mathbf{s}^\top\}],
\end{align}
where, $(\cdot)^\top$ denotes matrix transpose.} Also, we defined adjacency relation between both nodes with an adjacency matrix, $\mathbf{A}$, which is a $2\times2$ minor diagonal matrix (i.e., $K=2$). $\mathbf{A}$ defines a graph without any self-linked edge. In other words, only a simple graph is defined. This simple graph allows a relationship to be held in terms of phase information over the relation between real and imaginary attributes of the nodes. Moreover, it allows to \textcolor{black}{weigh} this information with an attention mechanism. The label input is also defined as: \textcolor{black}{$\mathbf{y} = [\mathbf{\theta};\, \mathbf{\varphi}]$}. 

In this study, $4$ consecutive \acp{GAL} are employed to learn features over the graph. In the sequel, we used a graph attention pooling to generalize the network and avoid it from overfitting. The proposed model also utilizes $4$ dense layers for the regression of the compressed signals over the angle pairs. \textcolor{black}{The proposed \ac{GAT} architecture for 2-D DoA is illustrated in~\FGR{fig:gat_scheme}.} \textcolor{black}{For nonlinearity,} ReLU is chosen as \textcolor{black}{an} activation function in each layer. \ac{MAE} is used as a loss function in this network since it provides a suitable metric for use in DoA estimation problems by its nature. As RMSprop has been proposed for fast and robust optimization~\cite{goodfellow2016deep}, it was preferred for the optimization method of the model in this study. The learning rate of \textcolor{black}{the} optimizer has been set to $5\times10^{\minus6}$. It is worth noting that performance evaluation of other optimization methods for the problem addressed by this study is an open issue for future studies.

Applying the procedure mentioned below, we created the training and test datasets for varying numbers of antenna masks. Hence, it should be noted that the number of features of each node, $F^{\prime}$, \textcolor{black}{becomes} $M$. For the training dataset, compressed signals with both $50$ dB SNR and 2-D DoA values (i.e., azimuth and elevation angles), which are uniformly distributed between $\minus30$\textdegree~and $30$\textdegree~with $1$\textdegree~steps, were prepared using varying numbers of antenna masks. \textcolor{black}{To ensure that the number of received signal samples for each azimuth and elevation angle pair is sufficient for learning, we chose the number of sample compressed signals as $50000$. In other words, for training, the number of signal samples for each angle pair is above 10}. The dataset is publicly available in~\cite{dataset}.

The network is trained with the data detailed above through $500$ epochs unless early stopping is activated. Non-decreasing loss for $20$ epochs evokes early stopping in this configuration. Thus, early stopping, besides $L_2$-regularization, prevents the network from overfitting.

\section{Numerical Results}\label{sec:result}

\begin{figure}[!t]
\centering
\subfigure[]{
\label{fig:mae}
\includegraphics[width=0.48\linewidth]{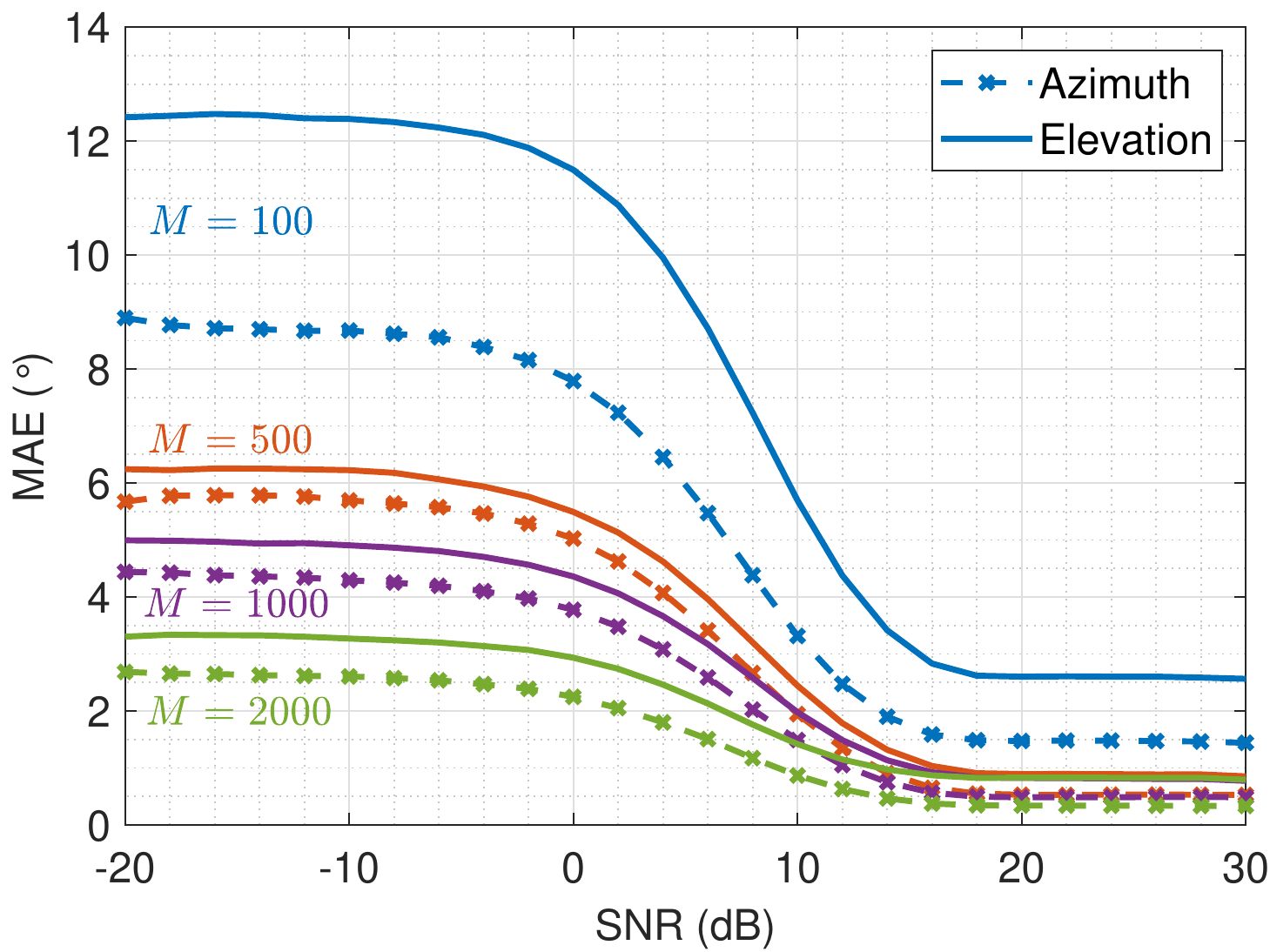}}
\subfigure[]{
\label{fig:comparison_mae_az_el_1000}
\includegraphics[width=0.48\linewidth]{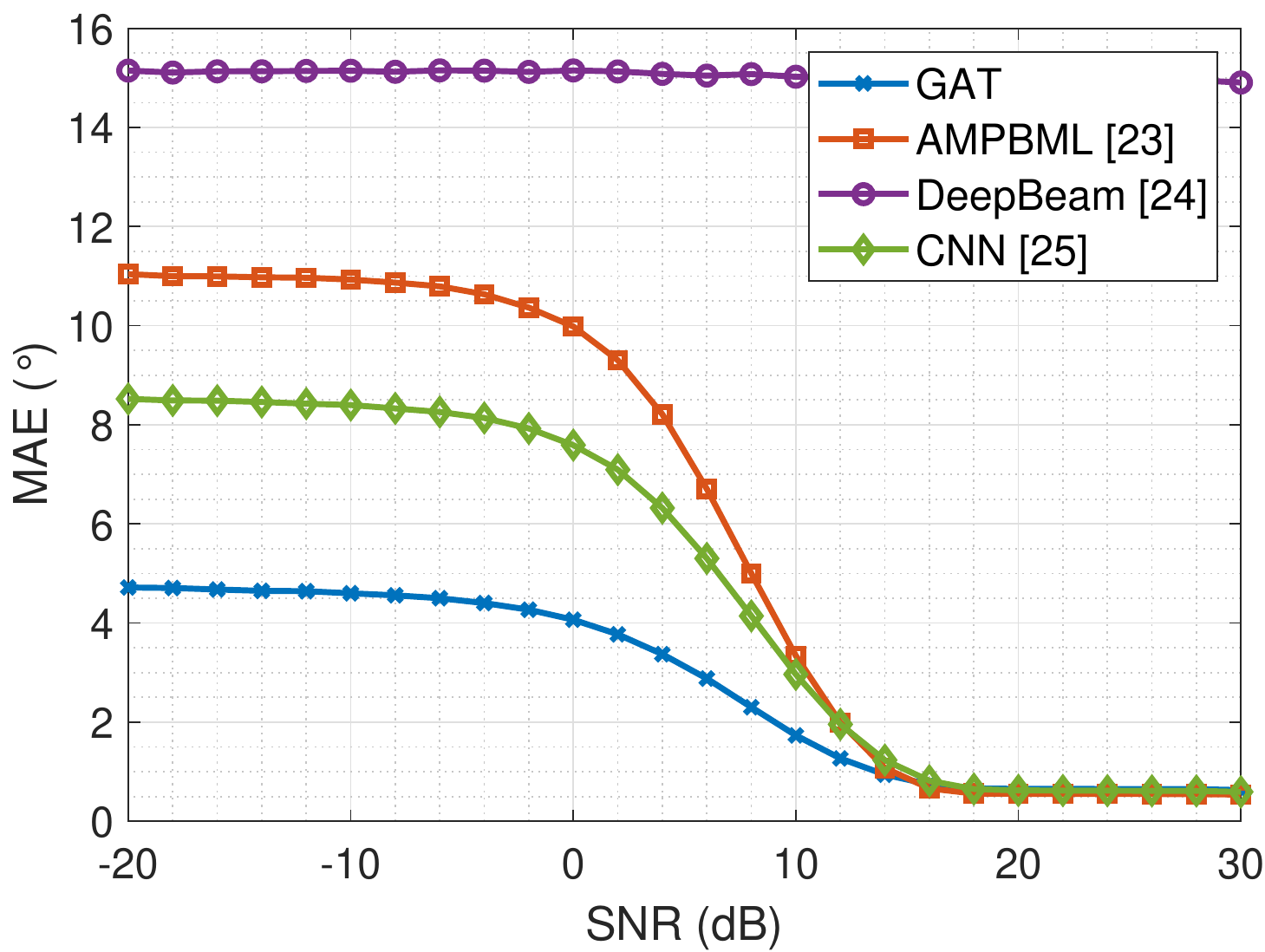}}
\caption{\textcolor{black}{MAE for (a) GAT-based 2-D DoA estimation with respect to SNR for varying numbers of antenna masks, where dashed and solid lines with the same color represent the performance for azimuth and elevation angle pairs, $\theta$ and $\varphi$, (b) joint azimuth and elevation estimation by various conventional DL methods and GAT when $1000$ masks are used.}}
\label{fig:comparison_mae_az_el_}
\end{figure}

The performance evaluation of the trained model is investigated for \textcolor{black}{a} various number of antenna masks. For performance evaluation, a test dataset was created at varying SNR levels and number of antenna masks. The test dataset consists of $5000$ single-pixel compressed signals received by metasurface aperture at each SNR level that is between $\minus20$ and $30$ dB. It should be noted that despite using signals at a high SNR level (i.e. $50$ dB) during training, the test process has been carried out under more challenging conditions (i.e. lower SNR). As detailed in \SEC{sec:method}, test data was also transformed to graph representation by defining the nodes and adjacency relation. $5000$ signals almost guarantee that one of each pair of angles is in the set. Thus, the results show an average performance for whole spatial scope. 

The initial results regarding the proposed proof-of-concept method demonstrate robust and high performance without any knowledge of the antenna transfer function and reconstruction step. The results given in~\FGR{fig:mae} show that even when the number of antenna masks is limited to $100$, \textcolor{black}{GAT can estimate the elevation and azimuth DoAs of the source with a mean absolute error (MAE) of $2.5$\textdegree~and $1.4$\textdegree, respectively}. Increasing the number of antenna masks improves the estimation accuracy whereas the computational complexity also increases. \textcolor{black}{For example, GAT using more than $500$ masks achieves less than $2$\textdegree~MAE at $10$ dB SNR for both elevation and azimuth}. It is observed that the estimation performance cannot improve while SNR value is increasing beyond $10$ dB. The reason for this phenomena is that the spatial scope is sampled with $1$\textdegree~step. Spatial sampling being a limiting factor can be clearly seen in~\FGR{fig:mae}. It is known that the smallest \ac{MAE} value for $1$\textdegree~interval is $0.5$\textdegree.
\FGR{fig:mae} shows that the results converge to the optimum value when more than $10$ dB SNR is used and more than $100$ masks are used. Oversampling might improve the estimation accuracy of \ac{GAT}, however, the training process would require more time and higher computational capacity due to increased data size. 

As seen in the results, the elevation estimation is slightly less accurate than the azimuth. The phenomena behind this situation can be explained by the \textcolor{black}{mask's} antenna pattern diversity in azimuth and elevation. It can be said that the radiation patterns of the designed metasurface aperture are more diverse in azimuth. In other words, as random measurement matrices improve sensing performance~\cite{do2011fast}, it can be considered that antenna masks for compressive sensing include more randomness in azimuth. \textcolor{black}{Although the aperture size and shape are the same in the azimuth and elevation direction, the metamaterial elements in the metasurface aperture can have an asymmetric radiation pattern~\cite{smith2017analysis}.}

\textcolor{black}{Furthermore, the comparative results for azimuth and elevation estimations with $1000$ masks are provided in~\FGR{fig:comparison_mae_az_el_1000}. It is demonstrated that the GAT technique shows a higher performance than conventional DL methods at low SNR regimes because of its attention mechanism. Morevoer, the proposed method has lower training time-complexity. For example, while the duration per epoch for GAT is $\tau$, AMPBML~\cite{ma2020machine}, DeepBeam~\cite{polese2021deepbeam}, and CNN~\cite{wu2019deep} require $1.8\tau$, $19.6\tau$, and $3.25\tau$, respectively. Here, $\tau$ is the duration per epoch and depends on the computational capacity.}

\section{Concluding Remarks}\label{sec:conclusion}
We presented a single-pixel DoA estimation technique leveraging GATs to learn directly from the compressed channel measurements to retrieve the DoA information. It was shown that the developed technique can achieve DoA estimation from the compressed measurements without the necessity for a reconstruction step. As a particular example, we showed that, using $M=2000$ aperture mask configurations, at an SNR level of $10$ dB, the estimation error in the azimuth and elevation angles were below $2$\textdegree, confirming the potential of the presented technique for high-fidelity DoA estimation using a single-pixel receiver architecture and without the need for an additional reconstruction step. \textcolor{black}{It was presented that the proposed method improves the estimation performance in 2-D at low SNR regimes compared to conventional DL methods. As stated, the attention mechanism can leverage the node attributes. Therefore, it can be argued that GAT networks can perform satisfactorily under a scattering environment. This aspect will form the basis of our future studies.} 


\balance

\bibliographystyle{IEEEtran}
\bibliography{gat_metasurface_doa}


\balance
\end{document}